\documentclass[12pt]{article}

\usepackage{sbc-template}
\usepackage{graphicx}
\usepackage{url}
\usepackage[utf8]{inputenc}
\usepackage[english]{babel}

\title{The Power of Attention: Bridging Cognitive Load, Multimedia Learning, and AI}

\author{Herbert dos Santos Macedo\inst{1}\thanks{These authors contributed equally to this work.}, 
Italo Thiago Felix dos Santos\inst{1}\footnotemark[1], 
Edgard Luciano Oliveira da Silva\inst{1}\footnotemark[1]}

\address{Escola Superior de Tecnologia -- Universidade do Estado do Amazonas (EST/UEA)\\
  Av. Darcy Vargas, Manaus -- AM, 69050-020 -- Brazil\\
  \email{hdsmj.lic22@uea.edu.br, itfds.eng20@uea.edu.br, elsilva@uea.edu.br}
}

\begin{document} 

\maketitle

\begin{abstract}
This article addresses the intersection of various educational theories and their relationship with the education of computer science students, with a focus on the importance of understanding computational thinking and its application in education. The historical context and fundamental concepts of Cognitive Load Theory, Multimedia Learning, and Constructivism are explored, highlighting their underlying biological assumptions about human learning. It also examines how these theories can be integrated with the use of Artificial Intelligence (AI) in education, with a particular emphasis on the attention mechanisms and abstract learning present in AI models like Transformers. Lastly, the relevance of these theories and practices for computer education student training is discussed, emphasizing how the development of computational thinking can contribute to a more effective approach in teaching and learning.
\end{abstract}


\section{Introduction}
The field of education is in constant evolution, shaped by new pedagogical theories and technological advancements. In recent years, traditional educational frameworks have increasingly intersected with innovative tools such as Artificial Intelligence (AI). In the context of computer science teacher education, this intersection plays a pivotal role in preparing future educators. Data from the INEP Higher Education Census \cite{inep2021} in 2021 indicate a significant shift in Brazilian higher education, with approximately 77\% of undergraduate students opting for Distance Learning (DL) courses. Concurrently, interest in teacher education programs experienced a 4-percentage-point decrease, dropping from 19\% to 15\% of offered slots, although federal networks maintained a higher enrollment average of 26\%. These figures highlight the necessity of deeper analyses regarding educational policies and teaching models.

This structural shift is further evidenced by the distribution of student choices within these modalities. Among the students who entered higher education in 2021, 55\% chose bachelor’s degree programs, 30\% opted for technological courses, and only 15\% enrolled in teacher education programs \cite{govbr_ead2022}. This disparity underscores a critical challenge: while the demand for digital literacy and computational skills grows exponentially, the pipeline for training qualified educators to teach these subjects is narrowing. The dominance of Distance Learning, combined with the low preference for licensure degrees, suggests that traditional pedagogical models are struggling to attract new generations of teachers. Consequently, there is an urgent need to innovate within these remaining programs, ensuring that future educators are not just consumers of technology but architects of effective digital learning experiences capable of addressing the specific cognitive and practical demands of Computer Science education.

Foundational pedagogical theories, such as Cognitive Load Theory, the Cognitive Theory of Multimedia Learning, and Constructivism, serve as essential pillars for translating technological innovations like AI into relevant teaching strategies. This convergence offers an opportunity to train future computer educators as learning architects capable of integrating modern technology with solid pedagogical foundations. Crucially, human learning is rooted in biological processes where synaptic activity strengthens or weakens neural connections through repetition and engagement. Analogously, computational models like artificial neural networks reflect aspects of these biological mechanisms. 

Machine learning models, most notably the Transformer architecture \cite{vaswani2017}, employ attention mechanisms to capture complex relationships within data. While Transformers learn abstract representations through computational attention, pedagogical methodologies guide students to progressively construct knowledge through practice, discussion, and interaction. By bridging biological assumptions, cognitive theories, and AI mechanisms such as cross-attention, educators can design learning experiences that deepen conceptual understanding and facilitate the application of knowledge in real-world scenarios.

\section{Related Works}

Understanding the cognitive processes involved in learning requires examining classical pedagogical frameworks alongside modern technological integrations. Cognitive Load Theory (CLT), formulated by Sweller \cite{sweller1988, sweller2011}, highlights the capacity limits of working memory and underscores the need for instructional designs that minimize cognitive overload. Building on developmental perspectives, Piaget \cite{piaget1964} demonstrated that learning is an active process of knowledge construction through environmental interaction rather than passive information reception. Complementing this, Gardner's Theory of Multiple Intelligences \cite{gardner2011frames} emphasizes that learners possess diverse cognitive capacities, requiring varied instructional methodologies to engage different brain functions effectively.

In parallel with pedagogical foundations, recent literature explores the application of AI in educational settings. Parreira, Lehmann, and Oliveira \cite{parreira2021} evaluated teachers' perceptions of AI technologies, revealing a generally positive attitude toward first-generation innovations while emphasizing the need to cultivate cross-cutting skills such as active listening, problem-solving, and research-oriented teaching. Addressing personalization, Cardoso et al. \cite{cardoso2023} explored how machine learning algorithms adapt course content and proposed an AI-based virtual tutor leveraging language models. Furthermore, Holmes et al. \cite{holmes2019artificial} provide a comprehensive overview of the promises and implications of AI in education, highlighting the need for ethical and pedagogical alignment. These applications rely fundamentally on the Transformer architecture \cite{vaswani2017}, which revolutionized sequence processing by replacing recurrent networks with self-attention mechanisms, enabling more efficient training across broad contextual domains.

\section{Objectives and Methods}

\subsection{Objectives}
The main objective of this work is to evaluate the intersection between foundational pedagogical theories and Artificial Intelligence (AI) in the education of computer science students, discussing instructional methodologies and the benefits of integrating AI concepts into teacher training, with a specific focus on fostering computational thinking.

\subsection{Methods}
This study adopted a documentary and bibliographical research methodology to gather and analyze theoretical and empirical data. The investigation involved examining scientific articles, books, technical reports, and conference proceedings addressing computer science education, cognitive load, multimedia learning, constructivism, and machine learning models, with a primary focus on the Transformer architecture \cite{vaswani2017}. Sources were selected based on their technical relevance and contribution to the proposed educational convergence.

\section{Traditional Methods and Cognitive Considerations}

The traditional teaching model, characterized by one-sided lectures where the instructor acts as the primary knowledge authority and students remain passive listeners, often fails to foster active investigation. In the digital age, where information is abundant and accessible, the primary educational challenge shifts from locating information to discerning, analyzing, and applying knowledge meaningfully. When applied passively, traditional instruction can induce severe cognitive overload. According to Cognitive Load Theory \cite{sweller1988, sweller2011}, managing cognitive capacity requires understanding how instructional formats interact with human memory limitations.

\textbf{Intrinsic Load:} Intrinsic load refers to the inherent complexity of the information to be learned, determined primarily by the degree of element interactivity within the subject matter. In traditional teaching environments, presenting complex, highly interconnected concepts in an unsegmented or rapid linear sequence severely exacerbates intrinsic load. When students are exposed to an overwhelming stream of complex data without strategic pauses or scaffolding, working memory capacity is rapidly exhausted, leading to the loss of critical details and hindering conceptual integration.

\textbf{Extrinsic Load:} Extrinsic load represents the additional cognitive effort imposed by external factors and poorly designed instructional materials that do not directly contribute to learning. In conventional passive lectures, where students act as mere spectators, the continuous mental effort required to maintain focus amidst monotonous or poorly structured presentations increases extrinsic load. This unnecessary expenditure of cognitive resources causes mental fatigue, reduces attention spans, and limits the working memory capacity available for actual knowledge processing.

\subsection{Constructivism and Multimedia Learning}

Constructivism posits that knowledge is actively constructed by the learner through experience, an idea expanded by Papert's Constructionism \cite{massa2022}, which emphasizes learning through creating tangible artifacts. In modern educational environments, multimedia tools allow students to express and consolidate understanding by integrating text, images, audio, and video. The Cognitive Theory of Multimedia Learning, extensively detailed by Mayer \cite{mayer2021}, grounds this practice by explaining how dual processing channels (visual and verbal) interact. Presenting complementary visual and verbal information activates both cognitive pathways, facilitating deeper connections and meaningful retention. This alignment creates beneficial redundancy, where simultaneous visual and verbal representations reinforce understanding rather than causing cognitive distraction.

\subsection{The Role of Cognitive Principles in AI-Generated Content}

While Mayer’s principles provide the theoretical foundation for effective multimedia design, the advent of Generative AI introduces new variables into this equation. AI tools can rapidly generate visualizations, code snippets, and explanatory texts, potentially overwhelming the learner if not carefully curated. According to the \textit{Coherence Principle}, extraneous material should be removed from instructional materials to avoid cognitive overload \cite{mayer2021}. In an AI-assisted environment, educators must act as filters, ensuring that the generated content aligns strictly with learning objectives. 

Furthermore, the \textit{Modality Principle} suggests that people learn better from graphics and narration than from graphics and on-screen text. AI-powered tutoring systems can leverage this by providing spoken explanations alongside dynamic code visualizations, rather than dumping large blocks of text. However, this requires a sophisticated integration where the AI understands not just the syntax of the code, but the pedagogical intent behind the visualization. Without such alignment, the sheer volume of available information can paradoxically increase extraneous load, contradicting the goal of efficient learning. Thus, the challenge shifts from content creation to content curation and pedagogical alignment.

\subsection{Biological Assumptions of Human Learning}

Understanding the biological underpinnings of learning provides vital insights for optimizing instructional design and drawing parallels with artificial neural networks. As explored by Damasio \cite{damasio2012}, cognitive processes are deeply rooted in neural architecture and brain function, where learning manifests as structural and chemical modifications across neuronal networks.

\textbf{Brain Plasticity:} Brain plasticity refers to the nervous system's fundamental capacity to structurally and functionally reorganize itself in response to environmental experiences and learning. This adaptability occurs primarily at synaptic junctions, where the strength of connections between neurons is dynamically adjusted based on activity frequency. Plasticity enables the brain to form new neural pathways, adapt existing circuits, and physicalize newly acquired knowledge and skills over time.

\textbf{Neurotransmitters and Neuronal Signaling:} Neuronal signaling relies on chemical messengers known as neurotransmitters to transmit electrical signals across synaptic clefts. Specific neurotransmitters play targeted roles during learning: dopamine modulates motivation, reward processing, and goal-directed focus, whereas acetylcholine is essential for synaptic encoding, memory formation, and attentional control. The coordinated interplay of these chemical signals governs the initial encoding and long-term stabilization of memories.

\begin{figure}[ht]
  \centering
  % Substitua '0.7\textwidth' se quiser ajustar o tamanho da imagem
  \includegraphics[width=0.7\textwidth]{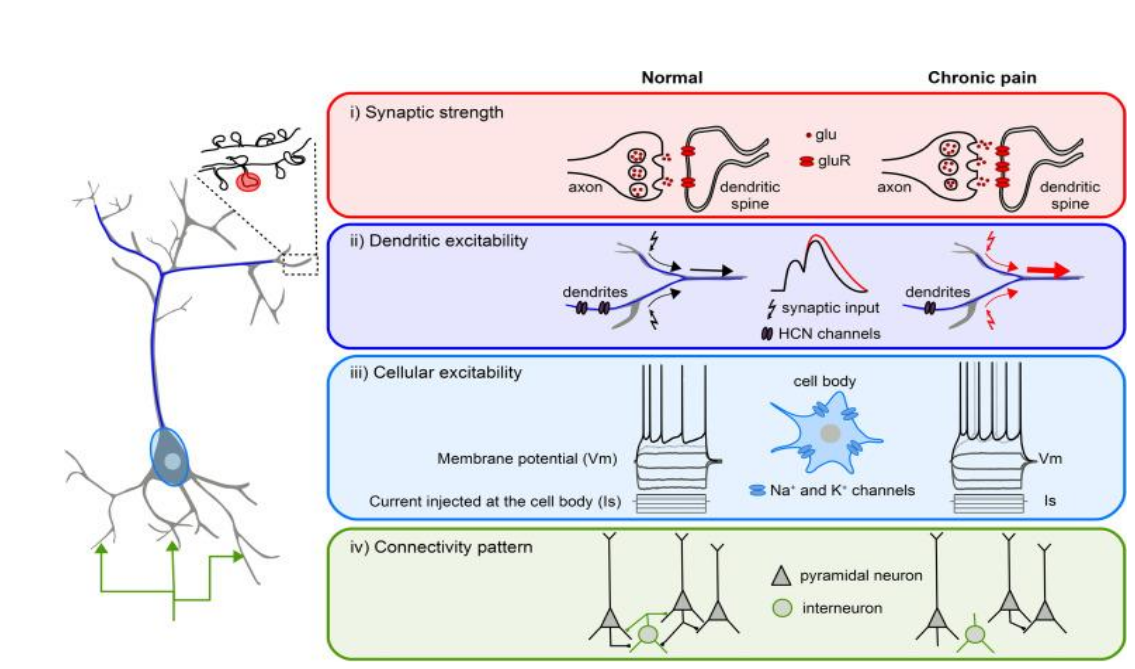} 
  \caption{Schematic Representation of Neuronal Adaptations \cite{kasanetz2022}. Note: While this study focuses on pathological plasticity, the underlying synaptic adaptation mechanisms are fundamental to general learning processes.}
  \label{fig:neuronal_adaptations}
\end{figure}

\textbf{Memory Consolidation:} Memory consolidation is the neurobiological process through which newly acquired, fragile short-term memory traces are progressively stabilized into long-term memory structures. During sleep, particularly deep slow-wave sleep, high-frequency neural reactivation occurs across hippocampal and cortical networks. This nocturnal reactivation reinforces synaptic connections established during waking learning episodes, rendering memories resilient against interference and decay.

\textbf{Long-Term Plasticity:} Long-term plasticity (LTP) represents a persistent strengthening of synapses based on recent patterns of high-frequency stimulation. This physiological process enhances signal transmission efficiency between paired neurons through structural changes, including increased neurotransmitter release and the enlargement of postsynaptic dendritic spines. LTP constitutes one of the primary cellular mechanisms underlying memory formation and persistent learning.

\textbf{Memory Spaces:} The human brain avoids storing information in centralized, isolated files, opting instead for distributed patterns of neural activation across specialized cortical regions. These interconnected networks form functional memory spaces, where distinct sensory, emotional, and conceptual aspects of a single experience are stored in parallel. Recalling a memory involves the synchronized reactivation of these distributed neural ensembles, enabling associative reasoning.

\textbf{Mechanisms of Emotion and Attention:} Emotion and attention operate as interconnected gatekeepers in the learning process. Emotionally salient stimuli engage subcortical structures like the amygdala, triggering the emotional processing effect that prioritizes memory encoding for impactful events. Concurrently, attention directs finite cognitive resources toward relevant environmental inputs, ensuring that prioritized concepts receive the deep processing required for retention.

\begin{figure}[ht]
  \centering
  % Ajuste '0.7\textwidth' conforme necessário para o tamanho da sua imagem
  \includegraphics[width=0.7\textwidth]{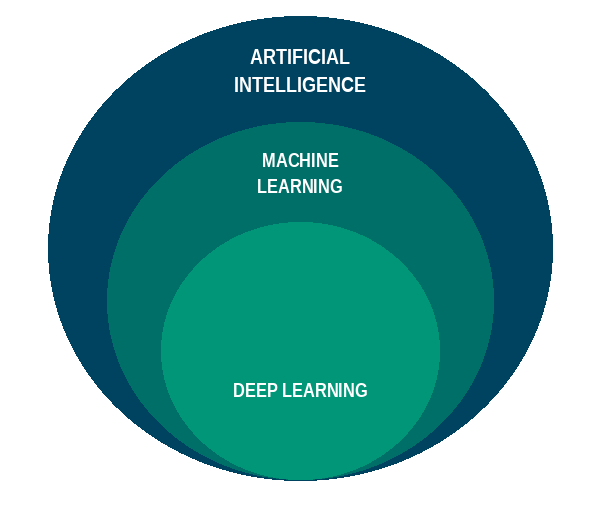}
  \caption{Hierarchy of Artificial Intelligence Types.}
  \label{fig:ai_hierarchy}
\end{figure}

These biological principles find computational counterparts in modern artificial intelligence models, particularly the Transformer architecture \cite{vaswani2017}. The Transformer utilizes self-attention mechanisms to dynamically weigh the importance of different input tokens within a sequence, capturing contextual relationships across broad sequence lengths.

\textbf{Distributed Attention:} Parallel to how human cognition allocates attention across multi-sensory inputs, distributed attention in artificial neural networks allocates computational weight to distinct elements across input sequences. In classroom environments, instructional strategies incorporating varied interactive media exploit distributed attention by guiding student focus across complementary aspects of the learning material.

\textbf{Emotional Activation and Selective Focus:} In biological learning, emotional resonance highlights priority information for deep encoding. Similarly, selective attention mechanisms within Transformer models assign higher numerical weights to critical tokens, ensuring that key contextual relationships dominate the model's internal representations during sequence generation.

\begin{figure}[ht]
  \centering
  % Ajuste '0.7\textwidth' conforme necessário para o tamanho da sua imagem
  \includegraphics[width=0.7\textwidth]{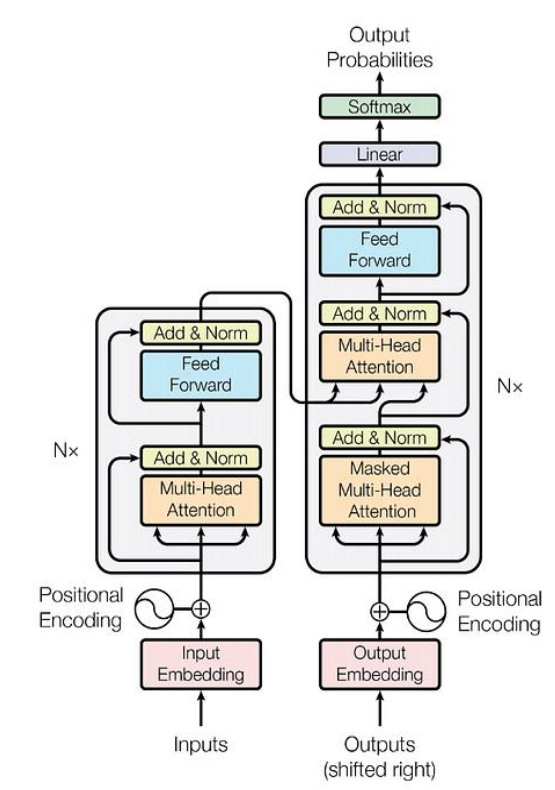}
  \caption{The Transformer - Model Architecture \cite{vaswani2017}.}
  \label{fig:transformer}
\end{figure}

Understanding these computational mechanisms allows educators to effectively integrate advanced AI tools, such as those developed by OpenAI \cite{openai2023}, into instructional environments to support diverse learning requirements.

\textbf{Personalization of Learning:} AI architectures can analyze student interaction patterns, error frequencies, and learning speed to dynamically adjust instructional difficulty and content presentation. Similar to how attention mechanisms emphasize relevant tokens, personalized AI environments tailor educational pathways to individual student profiles, optimizing engagement and conceptual mastery.

\textbf{Real-Time Feedback:} Language models can evaluate student inputs and generate immediate, context-aware feedback during writing, coding, or problem-solving tasks. Receiving immediate formative corrections allows students to identify and rectify misconceptions while the cognitive context remains active in working memory, accelerating the learning cycle.

\textbf{Support for Educators:} Beyond direct student interaction, AI models assist educators by automating routine tasks such as generating exercise sets, drafting lesson plans, and synthesizing assessment metrics. By reducing administrative workload, AI tools enable teachers to dedicate more time to high-value interpersonal interactions and targeted pedagogical support.

\subsection{Computational Thinking and AI Integration}

A core objective in computer science education is the development of \textbf{Computational Thinking (CT)}. Defined by Wing \cite{wing2006}, CT involves solving problems, designing systems, and understanding human behavior by drawing on concepts fundamental to computer science. Brennan and Resnick \cite{brennan2013} further expand this framework, identifying three dimensions of CT: concepts (such as abstraction and decomposition), practices (such as iterative testing and debugging), and perspectives (such as empathizing with users and celebrating creativity).

The integration of AI, particularly Large Language Models (LLMs), significantly supports the cultivation of these CT skills. By offloading rote syntactic errors and boilerplate code generation to AI assistants, students can redirect their limited working memory—protected from extraneous cognitive load—toward higher-order CT activities. For instance, AI can assist in \textit{decomposition} by suggesting modular structures for complex problems, allowing students to focus on the logical flow and algorithmic design. Furthermore, AI-driven simulations and debuggers provide immediate feedback loops that align with the iterative nature of programming, reinforcing the \textit{debugging} and \textit{pattern recognition} aspects of CT. Thus, AI serves not as a replacement for critical thought, but as a scaffold that elevates the cognitive level of student engagement with computational problems.

\subsection{Results and Discussion}

Synthesizing pedagogical theories with Artificial Intelligence highlights complementary dimensions of student development. From a cognitive perspective, Sweller \cite{sweller1988, sweller2011} demonstrates working memory constraints that instructional design must accommodate, while Piaget \cite{piaget1964} and Gardner \cite{gardner2011frames}, as revised by Morais Sales \cite{morais2018}, emphasize that learning involves active assimilation and diverse intellectual capacities requiring targeted educational interventions. When instruction accounts for these varied cognitive dimensions, students move beyond rote memorization toward conceptual mastery.

From an empirical and technological standpoint, studies by Parreira, Lehmann, and Oliveira \cite{parreira2021} and Cardoso et al. \cite{cardoso2023} illustrate both the potential and the necessary cautions of AI integration. While educators perceive AI positively, Parreira et al. \cite{parreira2021} stress the importance of balancing technological tools with critical teaching competencies to avoid over-saturation. Meanwhile, Cardoso et al. \cite{cardoso2023} show how transformer-based virtual tutors can monitor progress, personalize content, and support both students and instructors. Additionally, Holmes et al. \cite{holmes2019artificial} caution that without proper pedagogical grounding, AI may exacerbate inequities or reduce critical thinking if used uncritically. These findings indicate that AI should function not as a substitute for pedagogy, but as an adaptive framework that enhances instructional monitoring and individual guidance.

Ultimately, successfully intersecting these domains depends on redefining the roles of the student, the teacher, and AI. Students must transition from passive receivers to autonomous, critical agents capable of managing technological tools and developing robust computational thinking. Teachers, rather than acting merely as information transmitters, evolve into learning managers and researchers who combine pedagogical tradition with technological innovation.

\subsection{Ethical Considerations and Equity in AI-Enhanced Education}

The integration of AI in education is not merely a technical challenge but an ethical one. Holmes et al. \cite{holmes2019artificial} emphasize that AI systems can perpetuate existing biases if training data is skewed or if algorithms lack transparency. In the Brazilian context, where digital inequality remains a significant barrier, there is a risk that AI-enhanced education may exacerbate disparities between well-resourced institutions and those with limited infrastructure. Students in remote areas, such as those in the Amazon region served by UEA, may face challenges related to internet connectivity and access to high-performance devices required for seamless AI interaction.

Moreover, the reliance on AI for feedback and assessment raises concerns about academic integrity and the development of critical thinking skills. If students become overly dependent on AI for problem-solving, they may fail to develop the deep conceptual understanding necessary for computational thinking. Therefore, it is crucial to implement AI not as a replacement for human judgment, but as a tool that augments teacher capabilities while maintaining rigorous academic standards. Educators must be trained to identify potential biases in AI outputs and to guide students in using these tools responsibly. This includes teaching students how to verify AI-generated information and understand the limitations of algorithmic decision-making. Addressing these ethical dimensions is essential for ensuring that AI serves as an equitable force in education, promoting inclusion rather than exclusion.

\section{Conclusion}

This work examined the intersection of foundational pedagogical theories, biological learning mechanisms, and Artificial Intelligence within computer science education. By exploring concepts such as cognitive load, constructivism, multimedia learning, and the attention mechanisms of Transformer architectures, the study underscores how AI can be leveraged to personalize instruction and optimize learning strategies. Specifically, we highlighted how AI can scaffold the development of Computational Thinking by reducing extraneous cognitive load and supporting iterative problem-solving. As educational technologies continue to evolve, future research should expand upon these theoretical connections through empirical studies involving both educators and students, measuring the impact of AI-assisted learning on specific CT competencies.

\bibliographystyle{sbc}
\bibliography{references} % Certifique-se de que este arquivo contém todas as chaves citadas

\end{document}